\documentclass[a4paper,twoside]{article}

\usepackage{epsfig}
\usepackage{subfigure}
\usepackage{calc}
\usepackage{amssymb}
\usepackage{amstext}
\usepackage{amsmath}
\usepackage{amsthm}
\usepackage{multicol}
\usepackage{pslatex}
\usepackage{SciTePress}
\usepackage[small]{caption}

\usepackage{multirow} 
\usepackage{booktabs} 
\usepackage{lineno}

\begin{document}

\title{A Dynamic Programming Solution to the\\Monotonic Path of Minimal Cost in a 3-Rows Matrix}

\author{\authorname{Marcelo Cicconet and Davi Geiger}
\affiliation{New York University}
\email{cicconet@gmail.com, geiger@cims.nyu.edu}
}

\keywords{dynamic programming, optimal path}

\abstract{We consider the problem of finding the path of minimal
cost going from left to right in a 3-rows matrix, starting at the third row, and not going downwards, where there's an additional cost related to not changing rows,
such that the higher the change in intensity
within the row, the higher the cost of not moving upwards.}

\onecolumn \maketitle \normalsize \vfill

\paragraph{Problem Statement}

\begin{figure}[t!]
\begin{center}
	\includegraphics[width=\linewidth]{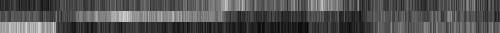}\\
	(a)\\
	\vspace{0.3cm}
	\includegraphics[width=\linewidth]{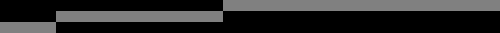}\\
	(b)
\end{center}
\caption{(a) Example of input. (b) Optimal path.}
\label{fig:dp}
\end{figure}

We consider a 3-rows matrix, similar to the one shown in Figure~\ref{fig:dp}~(a),
in which the values range from 0 to 1.
We are looking for the path of minimal
cost going from left to right, starting at the third row, and not going downwards.
That is, once the path is in row $i$, it can either remain
on that row or go to row $i-1$ (when $i-1$ exists).
Furthermore, there's an additional cost related to not changing rows,
such that the higher the change in intensity
within the row, the higher the cost of not moving upwards.

\paragraph{Solution by Dynamic Programming}
Dynamic programming involves building two matrices:
$Q$, of accumulated path costs, and $P$, of predecessors (which allows recovering
the path of minimal cost).

Let $C$ be the matrix of costs (e.g., Figure~\ref{fig:dp}~(a), with white color
representing cost zero).

We start by defining a ``windowed'' derivative, $D$.
Let $w$ be a window-size parameter; and $m$, $n$ the number of rows, columns in $C$,
respectively. For $j = w+1 \cdots n-w+1$, and $i = 1,2,3$,
set

$$D_{i,j} =  \frac{1}{w} \| C_{i,j \cdots j+w-1}-C_{i,j-w \cdots j-1} \|_1\text{ ,}$$
and the remaining columns of $D$ by replicating the closest filled column.

Let $\beta$ be a ``decay'' parameter. We define a ``derivative strength,'' $S$,
as

$$S(i,j) = \frac{1}{1+e^{-\beta D(i,j)}}\text{ .}$$
$S$ ranges from $1/2$ to $1$ when $D$ goes from $0$ to $\infty$.

The first column of $Q$ is set as the first column of $C$.
For $j = 2,\cdots,n$, we set

\vspace{-0.5cm}
\begin{eqnarray*}
s &=& \min \{ Q_{i,j-1}, Q_{i+1,j-1}+ \mu (1-S_{i+1,j}) \}\text{ ,}\\
s_a &=& \arg \min \{ Q_{i,j-1}, Q_{i+1,j-1}+ \mu (1-S_{i+1,j}) \}\text{ ,}\\
P_{i,j} &=& s_a+i-1 \text{ ,}\\
Q_{i,j} &=& s+C_{i,j}\text{ , for } i = 1,2\text{, end}\\
P_{3,j} &=& 3\text{ ,}\\
Q_{3,j} &=& Q_{3,j-1} + C_{3,j}\text{ ,}
\end{eqnarray*}

\noindent
where $s_a$ assumes values $1$ or $2$ and $\mu$ controls the
weight of the penalty $S$.
For the example shown in Figure~\ref{fig:dp} we used $w = 5$, $\beta = 7$,
and $\mu = 16$. These parameters are set empirically.

To recover the optimal path $p = \{p_1,...,p_n\}$, we compute
$p_n = \arg \min \{ Q_{1,n}, \cdots, Q_{m,n}\}$, and for
$j = n-1,\cdots,1$, we set $p_j = P_{p_{j+1},j+1}$.

Figure~\ref{fig:dp} shows an example of pair input/solution.

\paragraph{Reference}
Bellman, R. Dynamic Programming. Dover Publications, Incorporated. 2003.

\end{document}